\documentclass{egs}
\usepackage{amssymb}
\usepackage{times}
\usepackage[dvips,final]{graphicx}
\journal{\NPG}
\msnumber{3011}
\received{5 March 2003}
\accepted{2 June 2003}
\newcommand{\be}{\begin{equation}}
\newcommand{\ee}{\end{equation}}

\newcommand{\paref}[1]{(\ref{#1})}
\newcommand{\eqref}[1]{eq.\,(\ref{#1})}
\newcommand{\Eqref}[1]{Eq.\,(\ref{#1})}
\def\deg{\hbox{$^{\hbox{\small o}}$}}
\newcommand{\figwidth}{\hsize}    % In the final copy: 7.5cm or 16cm
\newcommand{\figref}[1]{{Figure~\ref{#1}}}
\newcommand{\tabref}[1]{Table~\ref{#1}}
\printfigures
\begin{document}
\title{Viscous heating in fluids with temperature-dependent 
viscosity: implications for magma flows}
\author[1]{A. Costa}
%\ead{costa@ov.ingv.it}
\affil[1,*]{Dipartimento di Scienze della Terra e Geologico-Ambientali,
Universit\`a di Bologna, Italy}
\affil[*]{also Dipartimento di Scienze della Terra, Universit\`a degli 
Studi di Pisa}
\author[2]{G. Macedonio}
%\ead{macedon@ov.ingv.it}
\affil[2]{Osservatorio Vesuviano, Istituto Nazionale di Geofisica e
Vulcanologia, Napoli, Italy.}%
%\author[cor]{Corresponding author. Address: G. Macedonio, Osservatorio
%Vesuviano - INGV, Via Diocleziano 328, I-80124 Napoli, Italy.}

%
%\thanks[now]{Also at Dipartimento di Scienze della Terra, Universit\`a
%degli Studi di Pisa, Italy.}
%
\maketitle
\begin{abstract}
Viscous heating plays an important role in the dynamics of fluids with
strongly temperature-dependent viscosity because of the coupling
between the energy and momentum equations.
The heat generated by viscous
friction produces a local temperature increase near the tube
walls with a consequent decrease of the viscosity which may
dramatically change the temperature and velocity  profiles.
These processes are mainly controlled by the Pecl\'et number, the
Nahme number, the flow rate and the thermal boundary conditions.
The problem of viscous heating in fluids was investigated in the past
for its practical interest in the polymer industry, and was invoked
to explain some rheological behaviours of silicate melts, but was not
completely applied to study magma flows.
In this paper we focus on the thermal and mechanical effects caused by
viscous heating in tubes of finite lengths.
We find that in magma flows at high Nahme number and typical flow
rates, viscous heating is responsible for the evolution from
Poiseuille flow, with a uniform temperature distribution at the inlet,
to a plug flow with a hotter layer near the walls.
When the temperature gradients induced by viscous heating
are very pronounced, local instabilities may occur and the triggering
of secondary flows is possible.
For completeness, this paper also describes magma flow in infinitely
long tubes both at steady state and in transient phase.
\end{abstract}
%\begin{keyword}
%Nonlinear fluids \sep non-isothermal rheology \sep Magma \sep Lava
%\end{keyword}
%
\section{Introduction}
A common feature of lava flows is the presence of a complex
network of lava tubes, observed both in ``pahoehoe'' \citep{pethol94}
and `a`a lava fields \citep{calpin98,calpin99}.
The tube system acts as an efficient
pathway for lava from the main vent to the flow fronts, allowing great
distances to be reached. This is mostly due to the decrease of
radiative cooling, as a consequence of a much lower temperature of the
outer crust of the tubes as compared to the temperature of the lava
surface flowing in open channels \citep{dra89,drapio95}.
\par
The effects of heat generation by viscous friction for lava
flows in tubes and channels were often neglected in previous models.
In this work we show that these effects can play an important role in
the dynamics of fluids with strongly temperature-dependent viscosity
such as silicate melts and polymers.
In fact, in these fluids, viscous friction generates a local increase
in temperature near the tube walls with consequent viscosity decrease
and increase of the flow velocity.
This velocity increase produces a further growth of the local
temperature.
As we will see later, there are some critical values of the parameters
that control this process above which this feedback cannot
converge. In this case the one-dimensional laminar solution, valid in
the limit of infinitely long pipe, cannot exist even for low Reynolds
numbers. 
In pipes with finite length, viscous heating governs the evolution
from Poiseuille regime with a uniform temperature distribution at the
conduit inlet, to a plug flow with a hotter boundary layer near the
walls downstream. This effect could be observed even in fluids with
null yield strength.
When the temperature gradients, induced by viscous heating are very
pronounced, local instabilities occur and triggering of secondary
flows is possible.
\par
Viscous heating was previously invoked to describe the rheological 
behaviour of basalts \citep{sha69} and to explain some
instabilities in the mantle
\citep{andper74,zhayue87,laryue95,hanyue96}. \\
Moreover \citet{fujuye74}, adopting a model only valid for infinitely long 
tubes (by the thermal point of view), explained the size of 
intrusive dikes in volcanoes. 
\par
Because of the low thermal conductivity of silicate melts, the
temperature field shows a strong radial gradient and viscosity
layering.
Flows with layers of different viscosity were investigated
in the past, also for their practical interest, and it is known that
they are not always stable \citep{yih67,cra69,renjos85,ren87}.
The instabilities are generated by the viscosity contrasts, and are
similar to the Kelvin-Helmholtz instabilities triggered by density
gradients.
In Couette flows of fluids with temperature-dependent
viscosity, \citet{suklau74} and \citet{yuewen96}
found one instability mode related to the viscosity gradient at low
Reynolds numbers.
This kind of instability shows a local character
\citep{pea76} and is expected to occur also in lava flows.
\par
As we will see later, these processes are controlled principally by
three parameters:
the Pecl\'et number ($Pe$), the Nahme number ($Na_0$, also called
Brinkman number), and the non-dimensional flow rate ($q$):
\begin{equation}
Pe=\rho c U H/k; \quad Na_0=\mu_0 U^2\beta/k; \quad 
q=\mu_0Q/(\rho gH^3) 
\end{equation}
with $\rho$ density, $c$ specific heat, $U$ mean velocity,
$H$ tube radius, $k$ thermal conductivity, $\mu_0$ reference
viscosity ($Na_0$ is based on this value), $\beta$ rheological
parameter (see eq.\ref{eq:viscosity_exponential}) and $Q$ flow rate
per unit length ($Q=UH$).
\par
Viscous heating in lava and magma flows is responsible for
effects not described by simple isothermal models, and may help in the
understanding of some phenomena observed in lava flows, which are not
yet clearly described.
These include the surprising temperature increase at the base of pahoehoe
flows observed by \citep{kes95b}, the formation of the ``roller vortex''
\citep{boosel73},
thermal erosion at low Reynolds numbers \citep{grefag98}, and
the observation, in lava channels, of magma temperatures greater than
the eruption temperature at the vent \citep{kaucas98}.
Erosion by viscous heating in mantle convection was also investigated
by \citet{laryue97} and \citet{moosch98}.
\par
Recently, the results of the studies of \citet{polpap2001} suggest that
the viscous heating may be responsible for the generation of a
heterogeneous distribution of magma properties inside the conduit.
Moreover, the recent viscous gravity currents model of
\citet{vasten2001}, based on the conservation equations applied to a
control volume, shows that the viscous dissipation exerts a strong
influence in the viscous gravity currents.
Finally, inadequacy of simple conductive cooling models is shown by 
basal temperature measurements carried out by \citet{kes95b}. 
Temperature measurements recorded at the base of many flows increases 
after some initial cooling. This fact can be easily understood on the 
light of the model presented here.
\par
We begin our study by investigating the limit case of flows in 
infinitely long tubes, and then we will describe the flow in tubes of 
finite length as being more representative of actual magma flows.
We find that viscous
heating effects cannot be generally neglected in typical magma flows
and that the energy and momentum equations cannot be decoupled.
Moreover, we find that the characteristic plug-like velocity profile
observed in magma flows can be caused by viscous heating effects,
even assuming Newtonian rheology.
\section{Model description}
\begin{figure}
\figbox*{}{}{%
\includegraphics[angle=0,width=\figwidth]{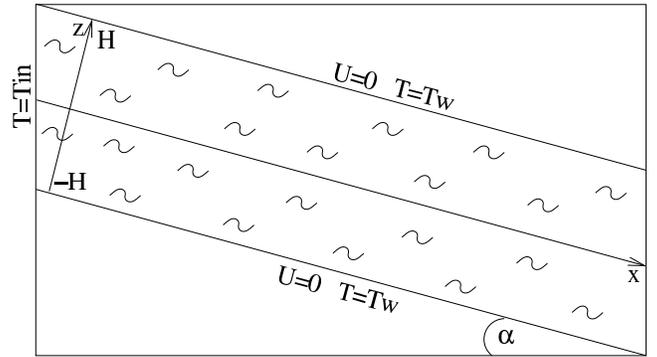}}
\caption{Sketch of the inclined lava tube and prescribed boundary
conditions.\label{fig:schema}}
\end{figure}

In this paper, magma is assumed to be incompressible and approximated as an
homogeneous fluid with constant density, specific heat and thermal
conductivity. The viscosity, however, is temperature-dependent.
We investigate the flow in a slab between
two parallel boundaries separated by a distance $2H$ and inclined of
an angle $\alpha$ with respect to the horizontal (see \figref{fig:schema}).
For symmetry, the velocity is directed along the flow.
At the tube walls ($z=\pm H$) we impose a null velocity
and a fixed temperature.
This last assumption on the temperature is quite restrictive, and
will be probably generalized in future works.
However, for the case of tube of finite length, we show also
results obtained by assuming thermal adiabatic condition at the walls,
as a limit case.
\par
When the Pecl\'et number is very high, the lubrication approximation is
valid: the transient term in the momentum equation is orders
of magnitude smaller than the corresponding term in the energy
equation (their ratio is of the order of the inverse of the Prandtl
number). This means that the time scale for momentum relaxation
is much shorter that the corresponding time scale for thermal
relaxation, and the time evolution of the velocity and temperature
profiles is controlled only by the energy equation.
In these hypotheses, magma dynamics in the tube are described by
the following transport equations:
\begin{eqnarray}
- \frac{\partial P}{\partial x}
+ \rho g \sin \alpha + \frac{\partial}{\partial z}\left
(\mu \frac{\partial U}{\partial z}\right)=0
\label{eq:momentum_dimensional}
\end{eqnarray}
and
\begin{eqnarray}
\rho c \left(\frac{\partial T}{\partial t}+U\frac{\partial T}{\partial
x}+W\frac{\partial T}{\partial z}\right) = k \frac{\partial^2
T}{\partial z^2} + \mu \left(\frac{\partial U}{\partial z}\right)^2
\label{eq:energy_general}
\end{eqnarray}
where $z$ is the coordinate transversal to the flow, $x$ longitudinal
coordinate, $t$ time, $(U,W)$ velocity of the fluid in the directions
$x$ and $z$ respectively, $\rho$ density, $P$ pressure, $\mu$ 
viscosity, $g$ gravity acceleration, $\alpha$ slope, $c$
 specific heat at constant pressure, $T$ temperature, and $k$ thermal
conductivity.
The last term in \paref{eq:energy_general} represents the heat
generation by viscous dissipation.
\par
The characteristic length scales of this problem are the channel dimensions
$2H$ (thickness) and $L$ (length), the mechanical relaxation length
$L_m=UH^2\rho/\mu_0$, and the thermal relaxation length $L_t=UH^2\rho
c/k$. 
For lava flows, typically $L_m/L\ll 1$, but $L_t/L\gg 1$ and the
approximation of infinitely long tube is not valid. Only when
$L_t/L\ll 1$, the approximation of infinitely long tube (by the
thermal point of view) is allowed.
This limit case is simpler to handle, and it was widely studied in the
past. For these reasons, we discuss magma flow in infinitely long
tubes, before studying the viscous heating effects in tubes of finite
length.
\par
In silicate melts the dependence of the viscosity on the temperature
is well described by the Arrhenius law
\citep{sha69,dan72,parive84,balpie86,cribal94}:
\begin{eqnarray}
\mu = \mu_A \exp(B/T)
\label{eq:viscosity_arrhenian}
\end{eqnarray}
where $\mu_A$ is a constant and $B$ is the activation energy. In the
interval of temperatures $(T-T_0)/T_0 \ll 1$ ($T_0$ is a
reference temperature), \eqref{eq:viscosity_arrhenian} is well approximated
by the (Nahme's) exponential law:
\begin{eqnarray}
\mu = \mu_0 \exp[-\beta(T-T_0)]
\label{eq:viscosity_exponential}
\end{eqnarray}
with $\beta = B/T_0^2$ and $\mu_0 = \mu_A \exp(B/T_0)$. In the
following, for simplicity, we adopt the exponential law, more
suited for the analytical manipulation of the equations
\citep{sha69,dra89,cosmac2002}.
\subsection{Infinitely long pipe}
In this section we study the flow in an infinitely long channel (slab)
driven only by the component of the gravity force acting along the
flow (all gradients are null along the flow).
We assume no-slip conditions ($U=0$),
constant temperature ($T=T_w$) at the walls ($z=\pm H$) and, for
simplicity, we choose the reference temperature $T_0=T_w$.
Under these hypotheses, \eqref{eq:momentum_dimensional} and
\paref{eq:energy_general} reduce to:
\be
\rho g \sin\alpha + \frac{d}{d z}\left[\mu(T)\frac{d U}{dz}\right]=0
\label{eq:momentum_steady}
\ee
and
\begin{eqnarray}
\rho c \frac{\partial T}{\partial t} =
k \frac{\partial^2 T}{\partial z^2} +
\mu \left(\frac{d U}{d z}\right)^2
\label{eq:energy_dimensional}
\end{eqnarray}
In a previous work, \citet{pea77} considered a Poiseuille flow
between two horizontal parallel planes with isothermal boundaries (the
temperature at the boundaries is the same as that of the injected
fluid), negligible transversal gradients, and a given flow rate.
Previously, a similar problem (the Hagen-Poiseuille flow)
was numerically solved using analogue computers by \citet{gruyou64}, 
and later \citet{eckfag86} studied the Couette flow.
Like \citet{pea77}, \citet{gruyou64} found that during the transient
phase, in contrast to the steady-state, the temperature profile shows
higher values near the walls than in the channel centre.
Moreover, \citet{gruyou64} found that the solutions of the equations
that they consider do not diverge only when the non-dimensional pressure
gradient  ${\mathcal G}_p$ $({\mathcal G}_p= \beta(dP/dx)^2
H^4/k\mu_0)$ is smaller than a critical value. In fact for 
${\mathcal  G}_p$ greater than this value, the temperature increases
indefinitely in a finite time. \citet{gruyou64} suggest that the
behaviour of this model can help in understanding the
origin of turbulence (however this feedback could not produce an
indefinite temperature increase in 2-D and 3-D systems).
\par
Next we investigate the time-dependent problem of a fluid with initial
temperature $T_i$, and wall temperature $T_w=T_0$, with $T_i \ge T_0$. We
study the conditions for the existence of the solution, and the
evolution to the steady-state solution (when it exists).
After combining \eqref{eq:momentum_steady} and
\paref{eq:energy_dimensional}, we rewrite the energy
equation in the non-dimensional form:
\be
\left\{
\begin{array}{lll}
\displaystyle
\frac{\partial \theta}{\partial \tau} = \frac{\partial^2 \theta}{\partial
\zeta^2} + {\mathcal G} \zeta^2 e^\theta \\
\displaystyle
\theta=0 & \forall \tau > 0, & \zeta = \pm 1 \\
\displaystyle
\theta = {\mathcal B} & \tau=0, & -1 < \zeta < 1\\
\end{array}
\right.
\label{eq:energy_adimensional_transient}
\ee
where
\be
\theta = \beta(T-T_0); \quad \zeta = \frac{z}{H}; \quad \tau=k
t/(\rho c H^2)
\ee
\be
\mathcal G = \frac{\beta(\rho g \sin \alpha)^2 H^4}{k \mu_0};
\quad \mathcal B = \beta(T_i-T_0)
\ee
The parameter $\mathcal G$ represents the non-dimensional ``shear
stress'', and is known as the Nahme number based
on the characteristic stress (instead of the velocity).
\par
\begin{figure}[htb!]
\figbox*{}{}{%
\includegraphics[angle=-90,width=\figwidth]{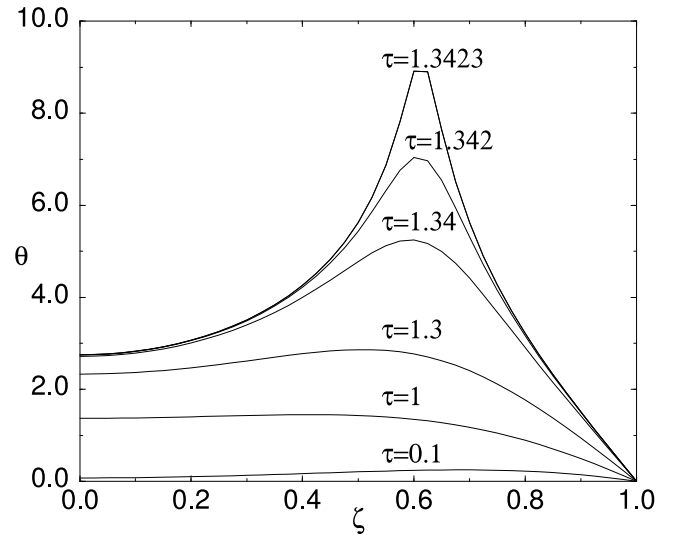}} 
\caption{Temporal evolution of the non-dimensional temperature profile in
the supercritical condition ${\mathcal G}=8$ and ${\mathcal B}=0$.
\label{fig:g8b0}}
\end{figure}
\begin{figure}[htb!]
\figbox*{}{}{%
\includegraphics[angle=-90,width=\figwidth]{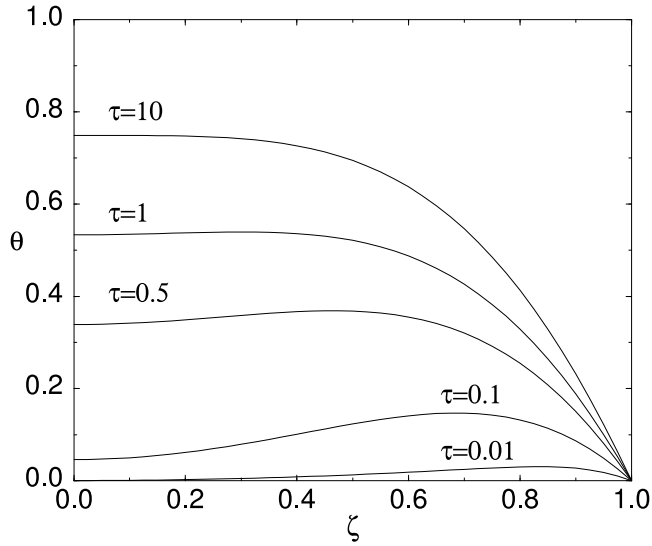}} 
\caption{Temporal evolution of the non-dimensional temperature profile in
the subcritical condition ${\mathcal G}=5$ and ${\mathcal B}=0$.
\label{fig:g5b0}}
\end{figure}

From the results of \citet{gruyou64} we know that when ${\mathcal B}=0$ and 
${\mathcal G}>{\mathcal G}_{crit}$ (${\mathcal G}_{crit} \approx 5.64$
for slab flows),
the solution of \eqref{eq:energy_adimensional_transient} diverges in a
finite time. In the case of ${\mathcal G}<{\mathcal G}_{crit}$,
the solution of the transient phase, with two maxima near
the walls, evolves for $\tau\approx 1$ to the steady-state solution
with only one maximum in the channel centre. In \figref{fig:g5b0}
and \figref{fig:g8b0}
we report the temporal evolution of $\theta$ found numerically
for the subcritical condition
(${\mathcal G}=5$), and for the supercritical
condition (${\mathcal G}=8$) respectively, both with ${\mathcal B}=0$.
As shown in \figref{fig:g5b0}, when ${\mathcal G}<{\mathcal G}_{crit}$,
for small $\tau$ the temperature 
increases near the boundaries and for $\tau \approx 1$ its profile tends
to the steady-state solution.
In \figref{fig:g8b0}, in the supercritical case, the temperature
profile, shows two peaks near the walls that continuously increase,
until they diverge when $\tau$ reaches a critical value $\tau_{crit}$.
\par
\begin{figure}[htb!]
\figbox*{}{}{%
\includegraphics[angle=-90,width=\figwidth]{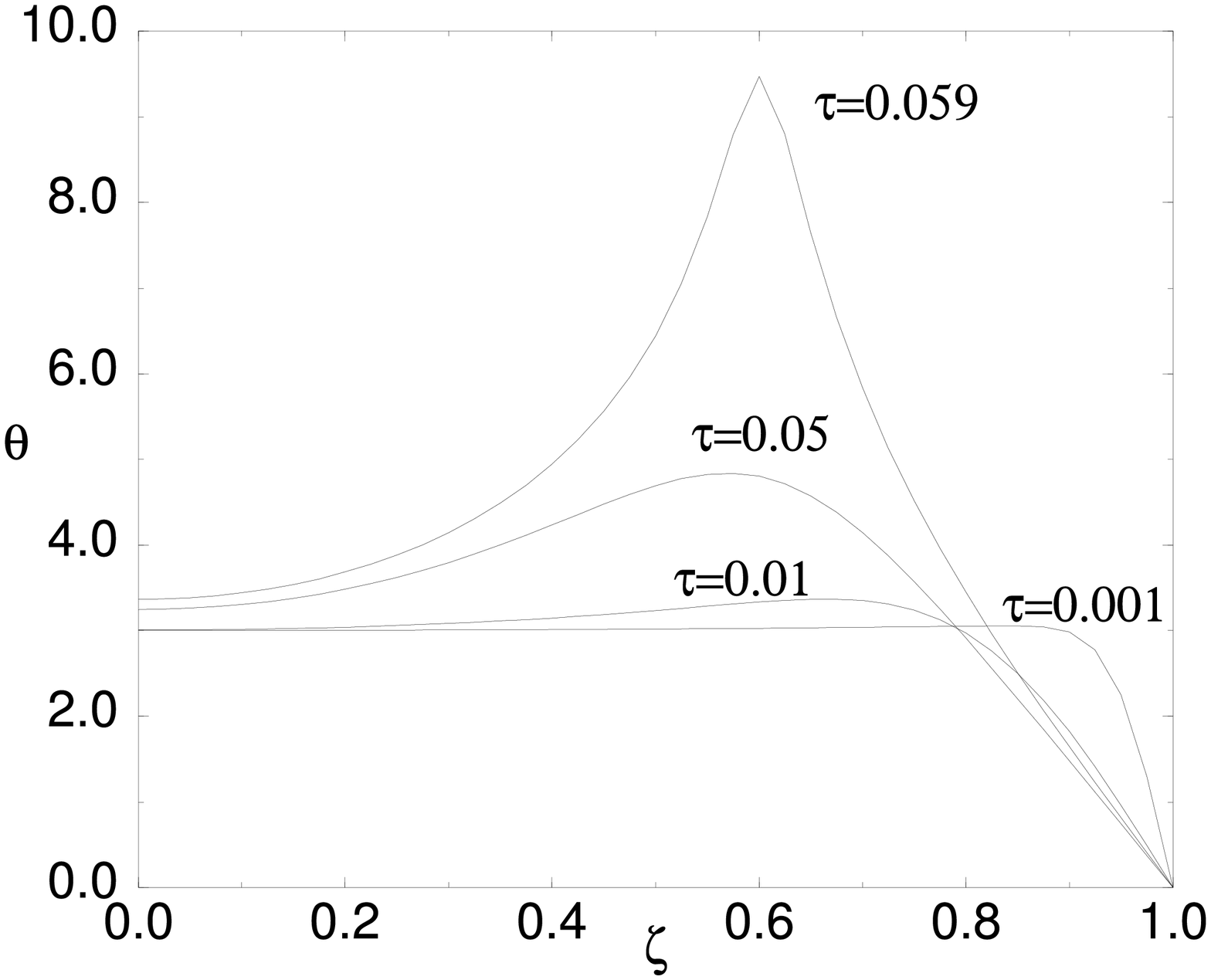}} 
\caption{Temporal evolution of the non-dimensional temperature profile in
the supercritical condition ${\mathcal G}=4$ and ${\mathcal B}=3$.
\label{fig:g4b3}}
\end{figure}
\begin{figure}
\figbox*{}{}{%
\includegraphics[angle=-90,width=\figwidth]{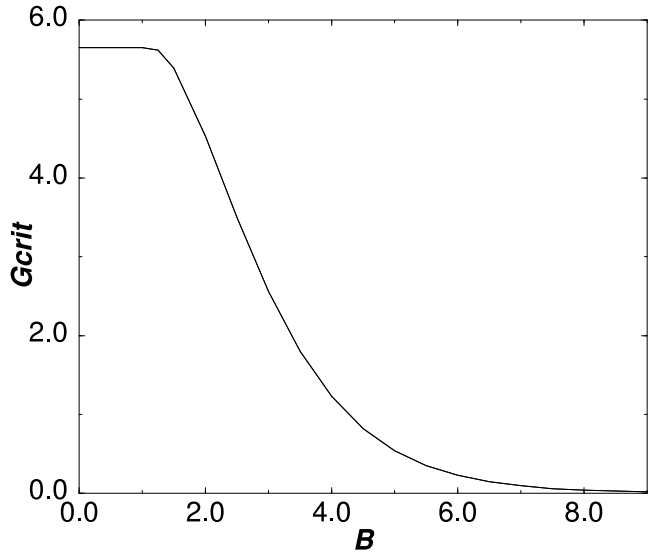}} 
\caption{Relation between ${\mathcal B}$ and ${\mathcal
G}_{crit}({\mathcal B})$.
\label{fig:gb}}
\end{figure}

The general case, with the initial fluid
temperature greater than the temperature at the walls
(${\mathcal B}>0$), was previously 
studied by \citet{fuj69} from a
theoretical point of view, but is not yet analytically or numerically
solved (to our knowledge).
\citet{fuj69} clarified the
relationship between the transient and the steady-state phases with
some theorems for the steady-state boundary value problem:
\be
\nabla^2 u +e^u = 0 \quad (x\in \Omega)
\label{eq:bvp_fujita}
\ee
and for the corresponding time-dependent problem:
\be
\frac{\partial u}{\partial t} = \nabla^2 u +e^u \quad (t\ge0, x\in \Omega)
\label{eq:ivp_fujita}
\ee
with initial condition $u(t=0)=a(x)$ (with $a(x)$ continuous), and
boundary condition $u(\partial \Omega)=0$.
One of the theorems states that if
\eqref{eq:bvp_fujita} has no solution, then the solutions of
\eqref{eq:ivp_fujita} diverge in a finite time, or diverge for $t
\rightarrow \infty$. Another theorem states that if
\eqref{eq:bvp_fujita} has more than one solution, $\phi$ is the
``smaller'' solution, and $\psi$ is another solution not equal to $\phi$,
then:
\begin{enumerate}
\item
If $a>\psi$, then the solution $u$ of \eqref{eq:ivp_fujita}
diverges, in a finite time, or for $t\rightarrow \infty$.
\item
If $a<\psi$, then the solution $u$ of \eqref{eq:ivp_fujita} uniformly
converges to $\phi$, for $t\rightarrow \infty$.
\end{enumerate}
From our numerical solutions of \eqref{eq:energy_adimensional_transient},
we observe that when ${\mathcal B}>0$, the range of values of
${\mathcal G}$ below which the solution of
\eqref{eq:energy_adimensional_transient} 
converges is smaller than the corresponding range of ${\mathcal G}$ for
${\mathcal B}=0$, but the profiles evolution is similar to the case
with ${\mathcal B}=0$.
We call ${\mathcal G}_{crit}({\mathcal B})$ the critical value of
${\mathcal G}$ above which the solution diverges when the initial
condition is ${\mathcal B}$. An example is reported in \figref{fig:g4b3}
for ${\mathcal G}=4$ and ${\mathcal B}=3$ where the temperature diverges
in a finite time.
\figref{fig:gb} shows the relation between ${\mathcal B}$ and
${\mathcal G}_{crit}({\mathcal B})$ found numerically.
\begin{figure}[htb!]
\figbox*{}{}{%
\includegraphics[angle=-90,width=\figwidth]{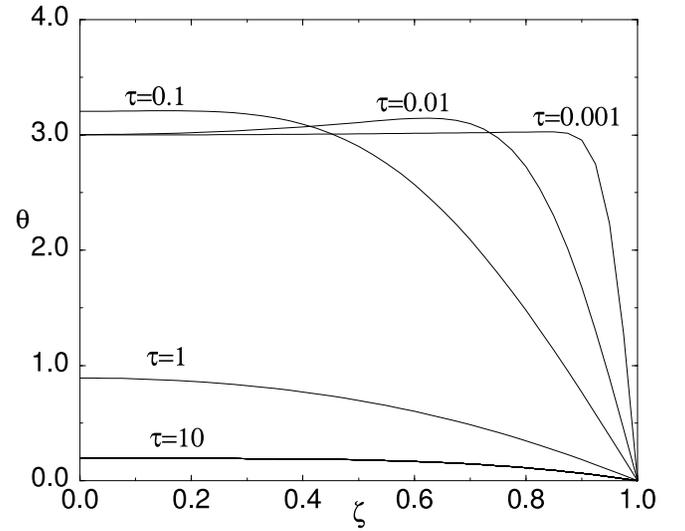}} 
\caption{Temporal evolution of the non-dimensional temperature for 
${\mathcal G}=2$ and ${\mathcal B}=3$.
\label{fig:g2b3}}
\end{figure}
For ${\mathcal B}>0$ we study the following two cases: in the
subcritical case (${\mathcal G}<{\mathcal G}_{crit}$) we investigate the
effect of the viscous heating on the temporal evolution of the temperature
profile, and we compare the results to the case without heat
generation (${\mathcal G}=0$); in the supercritical case, we
investigate the relation between the critical time
corresponding to the thermal blow-up $\tau_{crit}$ and ${\mathcal G}$.
When the viscous heating is neglected, 
\eqref{eq:energy_adimensional_transient} reduces to the classical heat
equation. In fact from an initial condition with uniform temperature
greater than that of the boundaries, the fluid starts cooling near the
walls, until the wall temperature is reached in a time scale of the
order of the characteristic time, i.e $\tau\sim 1 $. 
By accounting for the viscous heating, as shown, for example, in
\figref{fig:g2b3}, the temporal evolution of the temperature can be
described in four phases. The initial phase is characterized by the
cooling near the walls; in the second phase the viscous heating near
the walls produces the increase of the temperature above its initial
value and the temperature profile assumes two maxima in the external
part of the profile and a minimum in the central part. During the
third phase the two maxima migrate towards the central part of the
channel, and the profile is characterized by a maximum temperature in
the centre, that could be also greater than its initial value.
In the fourth phase the fluid cools until the steady state solution is 
reached, in agreement with the theoretical results of \citet{fuj69},
previously reported. 
\par
\begin{figure}[htb!]
\figbox*{}{}{%
\includegraphics[angle=-90,width=\figwidth]{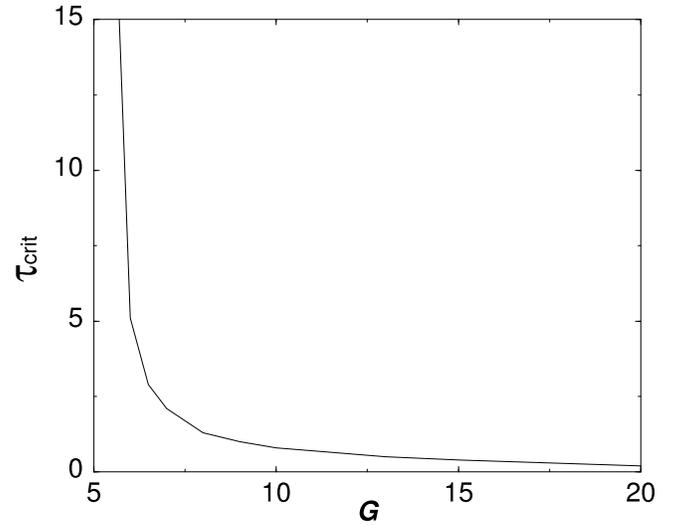}} 
\caption{Relation between $\tau_{crit}$ and ${\mathcal G}$ for
${\mathcal B}=0$.
\label{fig:taucrit}}
\end{figure}

In the supercritical case (${\mathcal G}>{\mathcal G}_{crit}$), we find an
non-dimensional critical time $\tau_{crit}$ above which the solution
diverges. In this case, the temporal evolution of the temperature
profile is qualitatively similar to the corresponding case with
${\mathcal B}=0$, but with shorter $\tau_{crit}$. The temperature
increases near the walls until it diverges as $\tau \rightarrow
\tau_{crit}$ (similar to the case shown in \figref{fig:g4b3}).
As ${\mathcal G}$ increases, $\tau_{crit}$ gets shorter.
As an example, \figref{fig:taucrit}
shows the relation between ${\mathcal G}$ and $\tau_{crit}$, for
${\mathcal B}=0$. 
\par
From the physical point of view, when viscous heating is relevant,
a transversal temperature gradient develops near the walls to
dissipate the heat through the boundaries. This temperature
increase produces a viscosity decrease with a
consequent increase of the flow velocity. This velocity increase
produces an increase in the velocity gradients near the walls, and a
further local temperature increase.
When ${\mathcal G}>{\mathcal G}_{crit}$, the temperature increases without
bound this feedback cannot converge and the steady-state solution
does not exist.
Of course, this has no physical meaning, and is related to the loss
of validity of some assumptions such as the hypothesis of
one-dimensional laminar flow. 
When the infinitely long pipe hypothesis is valid, according to
\citet{gruyou64}, it is legitimate to assume that above the critical
values (${\mathcal G}_{crit}$, $\tau_{crit}$), other diffusive
processes are triggered, allowing a more efficient heat and momentum
transfer.  
\par
\begin{figure}[htb!]
\figbox*{}{}{%
\includegraphics[angle=-90,width=\figwidth]{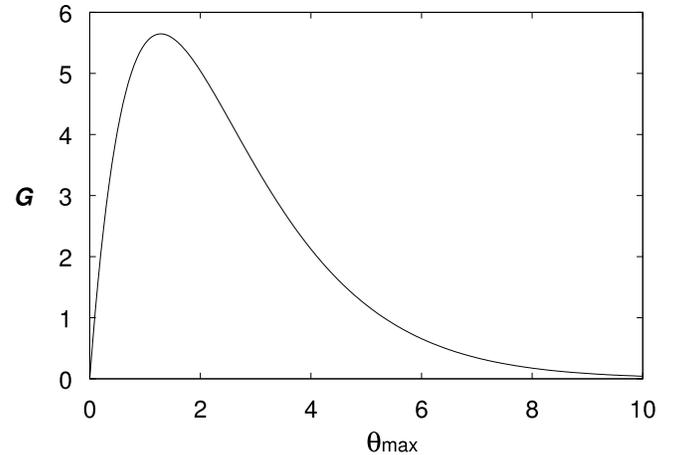}} 
\caption{Non-Dimensional ``shear-stress'' ${\mathcal G}$ vs. $\theta_{max}$.
($\theta_{max}$ is the maximum temperature in the channel centre, at
steady state.)
\label{fig:gcrit}}
\end{figure}

Instead, when $\mathcal G < \mathcal G_{crit}$, the temperature and
velocity profiles tend to the steady-state solutions described by:
\be
\frac{d^2\theta}{d\zeta^2} + {\mathcal G} e^\theta \zeta^2 = 0
\label{eq:theta_steady}
\ee
\Eqref{eq:theta_steady} was widely studied in the past, and it is
well known that it has no solutions for
${\mathcal G} > {\mathcal G}_{crit}$, where ${\mathcal G}_{crit}$ is a critical
value \citep{jos64}. Below the critical value, for each ${\mathcal G}$
there are two solutions corresponding to different temperature
profiles: one at low temperature, and one at high temperature (see
\figref{fig:gcrit}).
For some years, the existence of multiple solutions
corresponding to a given ``shear-stress'' $\mathcal G$ was discussed in the
literature, and the stability of the solution in the higher branch was
investigated. This depends on which is the controlling variable of the
problem: the velocity or the shear stress \citep{johnar97}. Moreover,
the relevance of the boundary conditions to the multiple steady-states
and the stability of the solution was investigated in the contest of
asthenospheric shear flow \citep{yuesch77}.
A complete review on this argument until 1974 is reported by
\citet{suklau74}, and more recently by \citet{becmck2000}.
In summary, in the Poiseuille flow between two parallel planes, when
the non-dimensional ``shear-stress'' parameter ${\mathcal G}$ is greater than
a critical value ${\mathcal G}_{crit}(\simeq 5.64)$,
\eqref{eq:theta_steady} has no solutions; for ${\mathcal G} =
{\mathcal G}_{crit}$ it has one solution;
and for ${\mathcal G} < {\mathcal G}_{crit}$
it has two solutions, one of which (the one with greater temperature) may
be unstable.
For ${\mathcal G} < {\mathcal G}_{crit}$, typical profiles
of the non-dimensional temperature $\theta$, and velocity $u/U_0$
that satisfy \paref{eq:momentum_steady}, \paref{eq:energy_dimensional}
and the imposed boundary conditions, are characterized by a maximum in 
the centre of the tube.
\subsection{Finite length tube}
From the thermal point of view, magma flows at high Pecl\'et
number cannot be described by assuming infinitely long tubes.
At high Pecl\'et numbers, the leading
term in the left side of \eqref{eq:energy_general} is the advective term,
containing the longitudinal gradient of the temperature
\cite[e.g.][]{lawkal97}.
Including only the relevant terms, the transport equations (mass,
momentum and energy balance) for finite length tubes, in non-dimensional
form are:
\begin{eqnarray}
\begin{array}{ll}
\displaystyle
\int_0^1 ud\zeta=q & \hbox{\small (mass)} \\
\displaystyle
\frac{\partial u}{\partial \zeta}=\left(\frac{\partial
p}{\partial\xi}-\sin\alpha\right)\zeta e^{\theta} &
\hbox{\small (momentum)}\\
\displaystyle
Pe^*u\frac{\partial
\theta}{\partial\xi} = \frac{\partial^2
\theta}{\partial \zeta^2} +Na^*\left(\frac{\partial
u}{\partial\zeta}\right)^2 e^{-\theta} & \hbox{\small (energy)}\\
\label{eq:adimensional_equations}
\end{array}
\end{eqnarray}
where $u=U/U^*$ with $U^* = \rho gH^2/\mu_0$, and $p$ non-dimensional pressure
($p=P/\rho gH$); moreover: 
\be
q=\frac{\mu_0Q}{\rho g H^3}
\label{q}
\ee
\be
Pe^*=\frac{\rho^2cgH^3}{k\mu_0}=\frac{Pe}{q}
\ee
\be
Na^*=\frac{\beta\rho^2 g^2 H^4}{k\mu_0}=\frac{Na_0}{q^2}
\label{eq:nahme_star}
\ee 
where $Na^*$ and $Pe^*$ are the Nahme and the Pecl\'et numbers based on
the characteristic velocity $U^*$ respectively and $Q=UH$.
The equation describing mass conservation is written in integral form.
To solve the equations in \paref{eq:adimensional_equations}, we
assume no-slip conditions for the velocity and constant temperature
$T_w$ at the walls. At the inlet we assume a parabolic velocity profile
and constant temperature $T_{in}$. Here, for
simplicity and according to other authors, we consider the wall
temperature as reference,
i.e. $T_0=T_w$ and $\mathcal B=\beta(T_{in}-T_w)$.
From an operative point of view, it would probably be more convenient
to set the reference temperature equal to the temperature at the inlet.
Of course, the physical behaviour of the flow is independent of the
choice of the reference temperature $T_0$; this is obtained by
rescaling all the parameters.
As an example, by assuming $T_0=T_{in}$, the viscosity
needs to be rescaled by a factor $e^{\mathcal B}$, etc.
\par
Recently, \citet{lawkal97} found analytical solutions of the simpler
problem with constant viscosity and showed typical temperature profiles
with two peaks near the walls. In our case, since we consider
temperature-dependent viscosity, we expect a feedback on the velocity
profile. 
In fact, using asymptotic solution method for high Nahme
numbers, \citet{ock79} solved the \eqref{eq:adimensional_equations}
taking in to consideration the full advective term. These solutions
describe the evolution of a flow from the Poiseuille regime with
uniform temperature at the conduit inlet to a plug regime
characterized by two peaks in the temperature profile near the
boundaries far from the inlet. 
Since the ratio $L_t/L_m$ (equal to the Prandtl number) is very high,
it is legitimate to assume that the flow is fully developed, by the
mechanical point of view, almost at the tube inlet.
The characteristic distance from the inlet where the viscous heating
becomes relevant, is $\xi^* \sim Pe Na^{-3/2}$ \citep{ock79} (using data 
from Table~\ref{tab:parametri_etna} this gives $50\div5000$m for a 
conduit width of 5m).
Under more general conditions, \citet{sch76}
found numerical solutions of the viscous heating problem, showing
clearly that the deviation from the Newtonian behaviour may have a
thermal origin in fluids with temperature-dependent viscosity.
When the viscous dissipation becomes important, its effect overcomes
the thermal cooling from the walls producing a maximum in the
temperature profile near the walls and a consequent plug-like velocity
distribution. These effects increase with distance from the inlet and,
unexpectedly, the maximum temperature gets more pronounced when
the temperatures at the walls get lower.
Similar results were obtained by \citet{galtak75} using perturbative
methods: they found that the flow rate is not proportional to the
longitudinal pressure gradient and the apparent viscosity of a
Newtonian fluid depends on the shear rate because of the viscous
dissipation. 
\par
We solve \eqref{eq:adimensional_equations} by using a
finite-difference method with an implicit scheme for the integration
along direction $\xi$; the pressure gradient is iteratively
adjusted at each step in order to satisfy mass conservation.
For symmetry we investigate only half a channel ($0\le \zeta \le 1$).
We find that the process is controlled by four parameters: a Pecl\'et
number $Pe$, a Nahme number $Na^*$, the non-dimensional flow rate
$q$, and the non-dimensional input temperature ${\mathcal B}$. However, since
we focus on magma flows in tubes, we set $Pe=10^7$
as a typical value, and we perform a parametric study on the others.
\par
\begin{figure}[htb!]
\figbox*{}{}{%
\includegraphics[angle=-90,width=\figwidth]{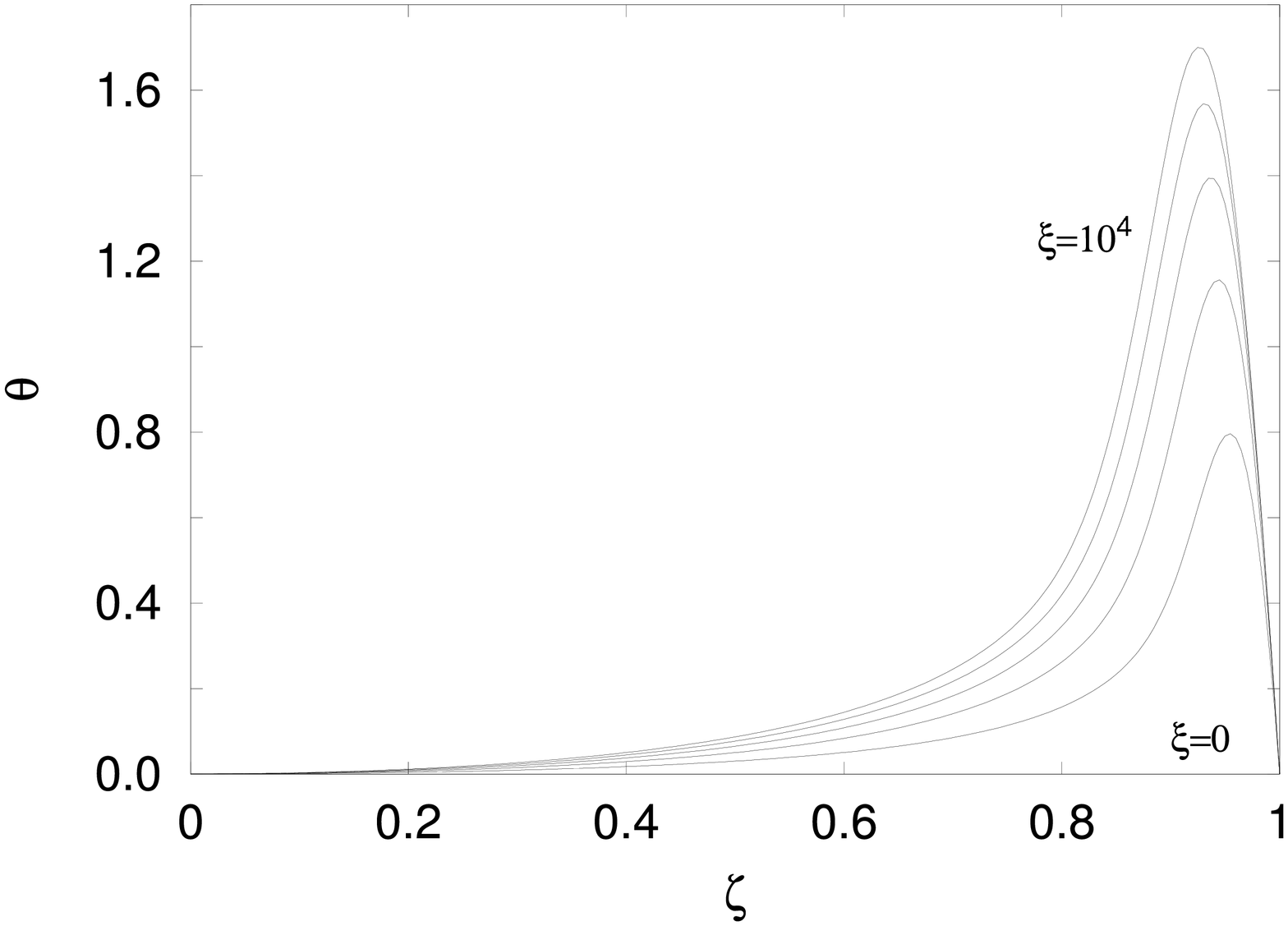}} 
\caption{Longitudinal evolution of the non-dimensional temperature
profile $Pe=10^7$, $Na^*=100$, $q=1$, ${\mathcal B}=0$ for different
non-dimensional distance from the inlet ($10^{-3}\xi = 0,2,4,6,8,10$).
\label{fig:temp-g100q1b0}}
\end{figure}
\begin{figure}[htb!]
\figbox*{}{}{%
\includegraphics[angle=-90,width=\figwidth]{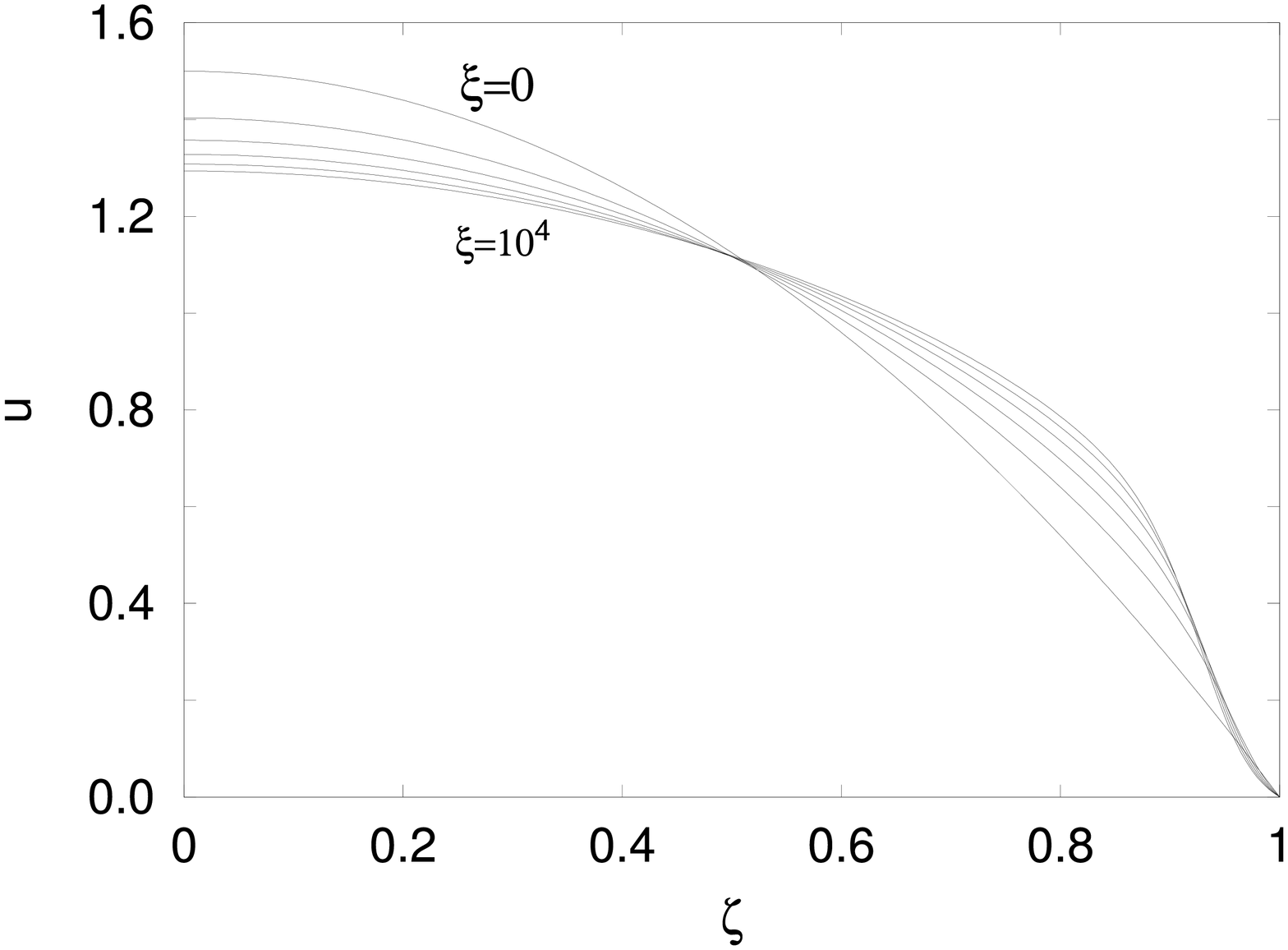}} 
\caption{Longitudinal evolution of the non-dimensional velocity profile
$Pe=10^7$, $Na^*=100$, $q=1$, ${\mathcal B}=0$ for different non-dimensional
distance from the inlet ($10^{-3}\xi = 0,2,4,6,8,10$).
\label{fig:vel-g100q1b0}}
\end{figure}
\begin{figure}[htb!]
\figbox*{}{}{%
\includegraphics[angle=-90,width=\figwidth]{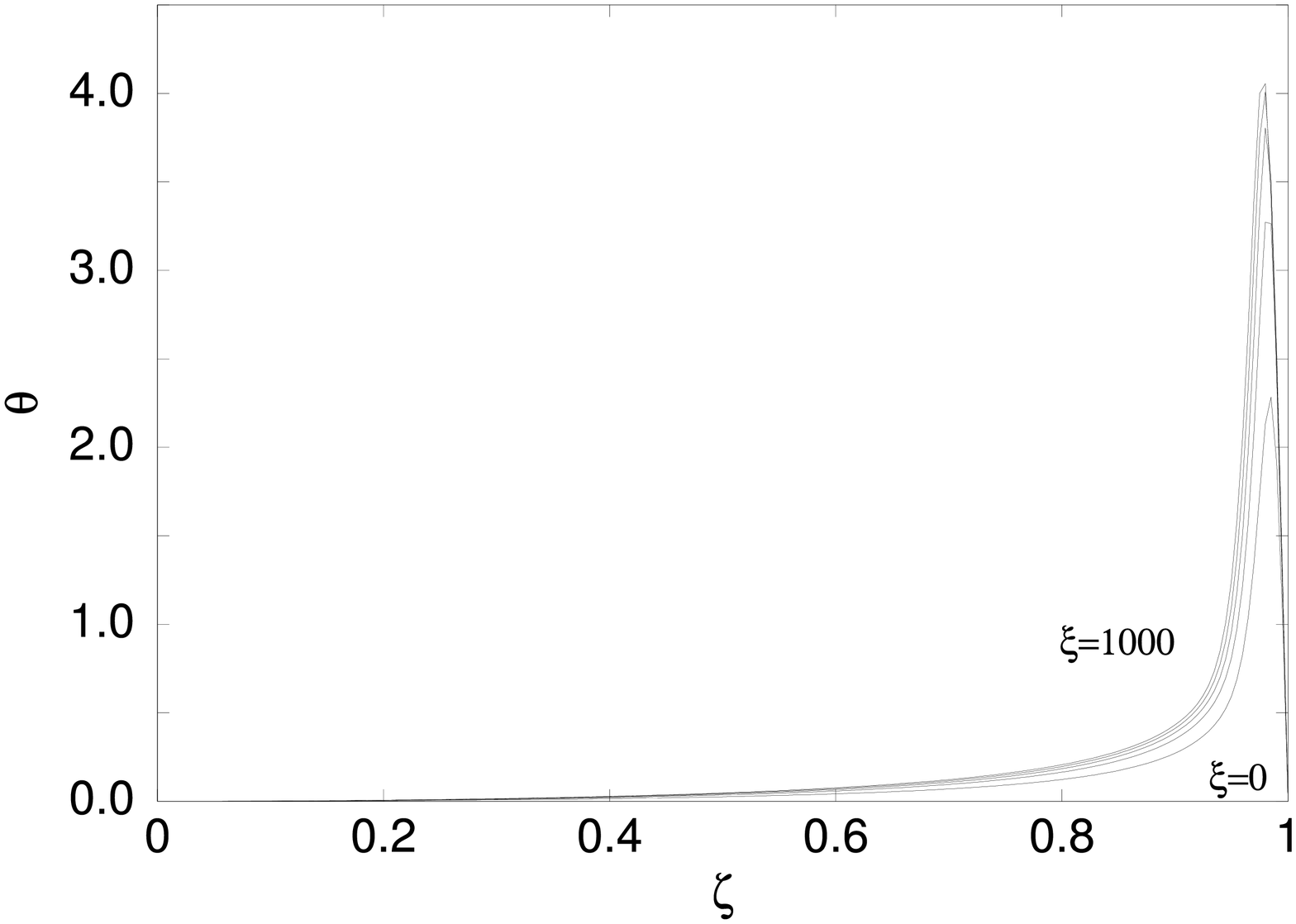}} 
\caption{Longitudinal evolution of the non-dimensional temperature
profile $Pe=10^7$, $Na^*=1000$, $q=1$, ${\mathcal B}=0$ for different
non-dimensional distance from the inlet ($\xi = 0,200,400,600,800,1000$).
\label{fig:temp-g1000q1}}
\end{figure}
\begin{figure}[htb!]
\figbox*{}{}{%
\includegraphics[angle=-90,width=\figwidth]{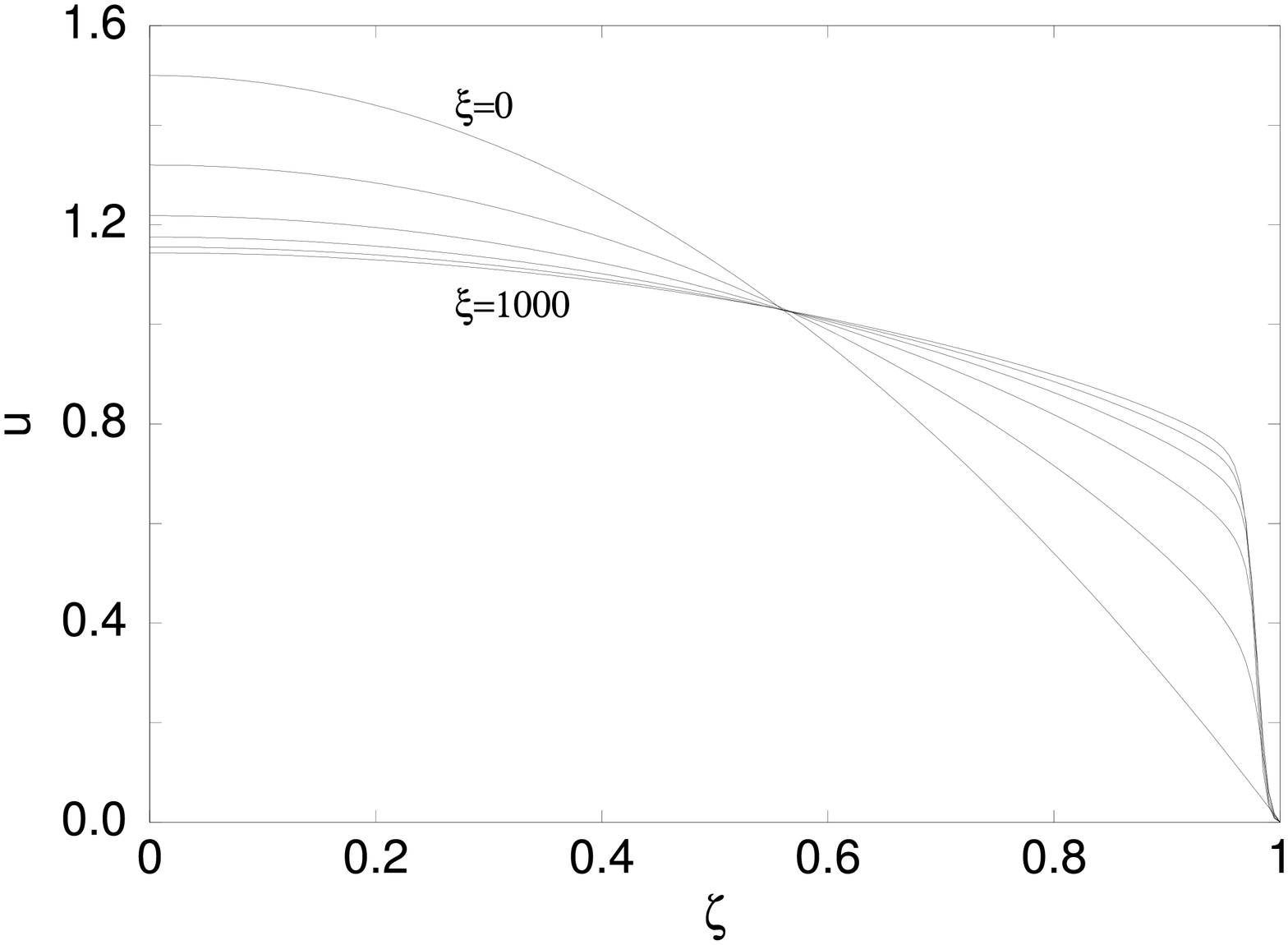}} 
\caption{Longitudinal evolution of the non-dimensional velocity profile
$Pe=10^7$, $Na^*=1000$, $q=1$, ${\mathcal B}=0$ for different non-dimensional
distance from the inlet ($\xi = 0,200,400,600,800,1000$).
\label{fig:vel-g1000q1}}
\end{figure}

In \figref{fig:temp-g100q1b0} and \ref{fig:vel-g100q1b0}, we show a
solution of \eqref{eq:adimensional_equations} for different non-dimensional
distance from the inlet: from $\xi=0$ to $\xi=10^4$.
Starting with uniform temperature ($\theta=0$) and parabolic velocity
profile at 
the inlet, the flow evolves gradually to a plug-like velocity profile
with two symmetric peaks in the temperature distribution.
In this case we used $Na^*=100$, $q=1$ and
${\mathcal B}=0$ and the viscous dissipation effects become important only
for high values of $\xi$. Instead in \figref{fig:temp-g1000q1} and
\ref{fig:vel-g1000q1} the solutions for $Na^*=1000$, $q=1$ and
${\mathcal B}=0$ are shown; these results are qualitatively similar to the
previous ones but the
temperature peaks are more pronounced and the length scale for the
development of the plug flow is lower than before (practically,
the transition occurs for $\xi=1000$).
\par
\begin{figure}[htb!]
\figbox*{}{}{%
\includegraphics[angle=-90,width=\figwidth]{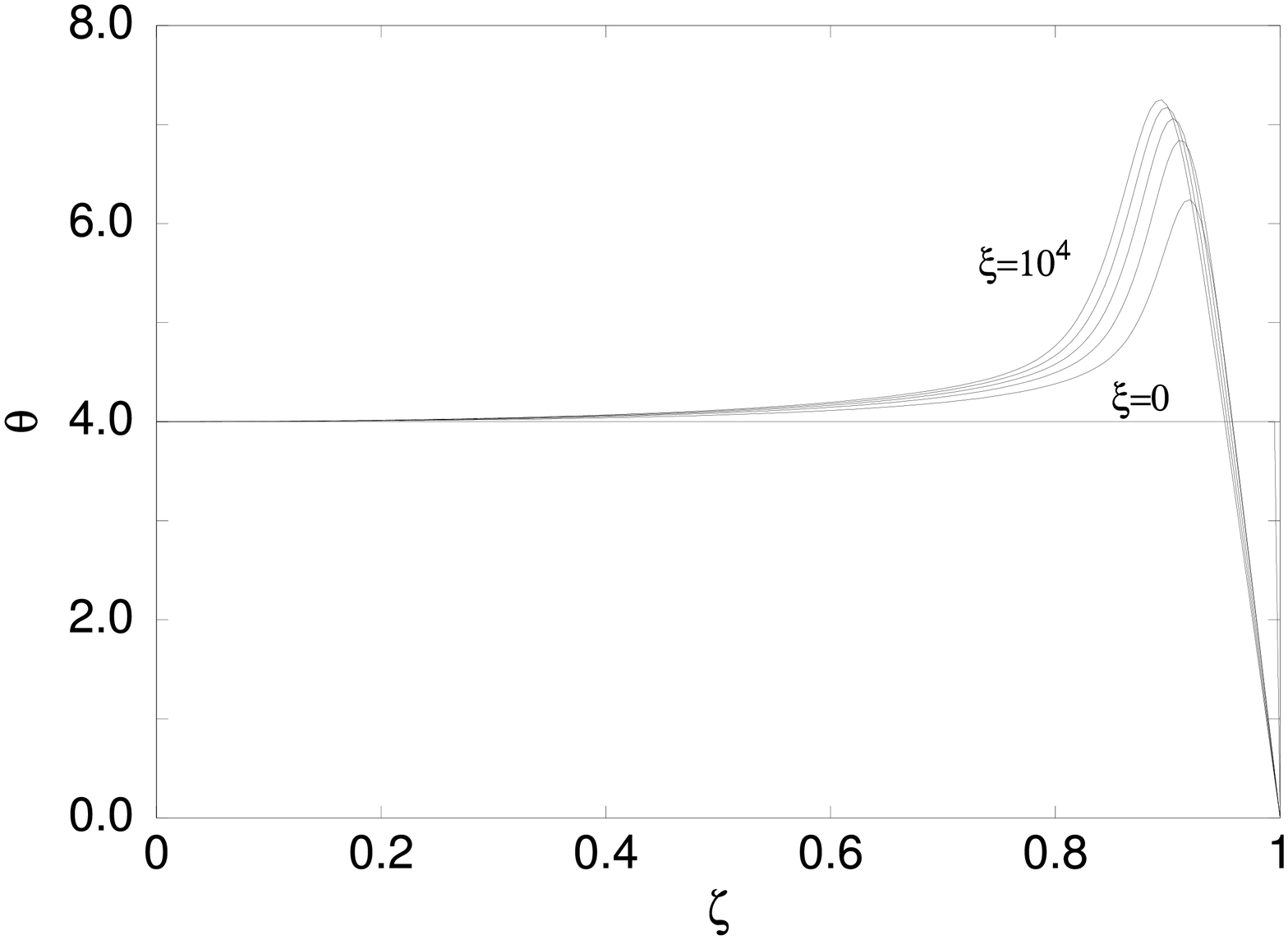}} 
\caption{Longitudinal evolution of the non-dimensional temperature
profile $Pe=10^7$, $Na^*=1000$, $q=4$, ${\mathcal B}=4$ for different
non-dimensional distance from the inlet ($10^{-3}\xi = 0,2,4,6,8,10$).
\label{fig:temp-g1000q4b4}}
\end{figure}
\begin{figure}[htb!]
\figbox*{}{}{%
\includegraphics[angle=-90,width=\figwidth]{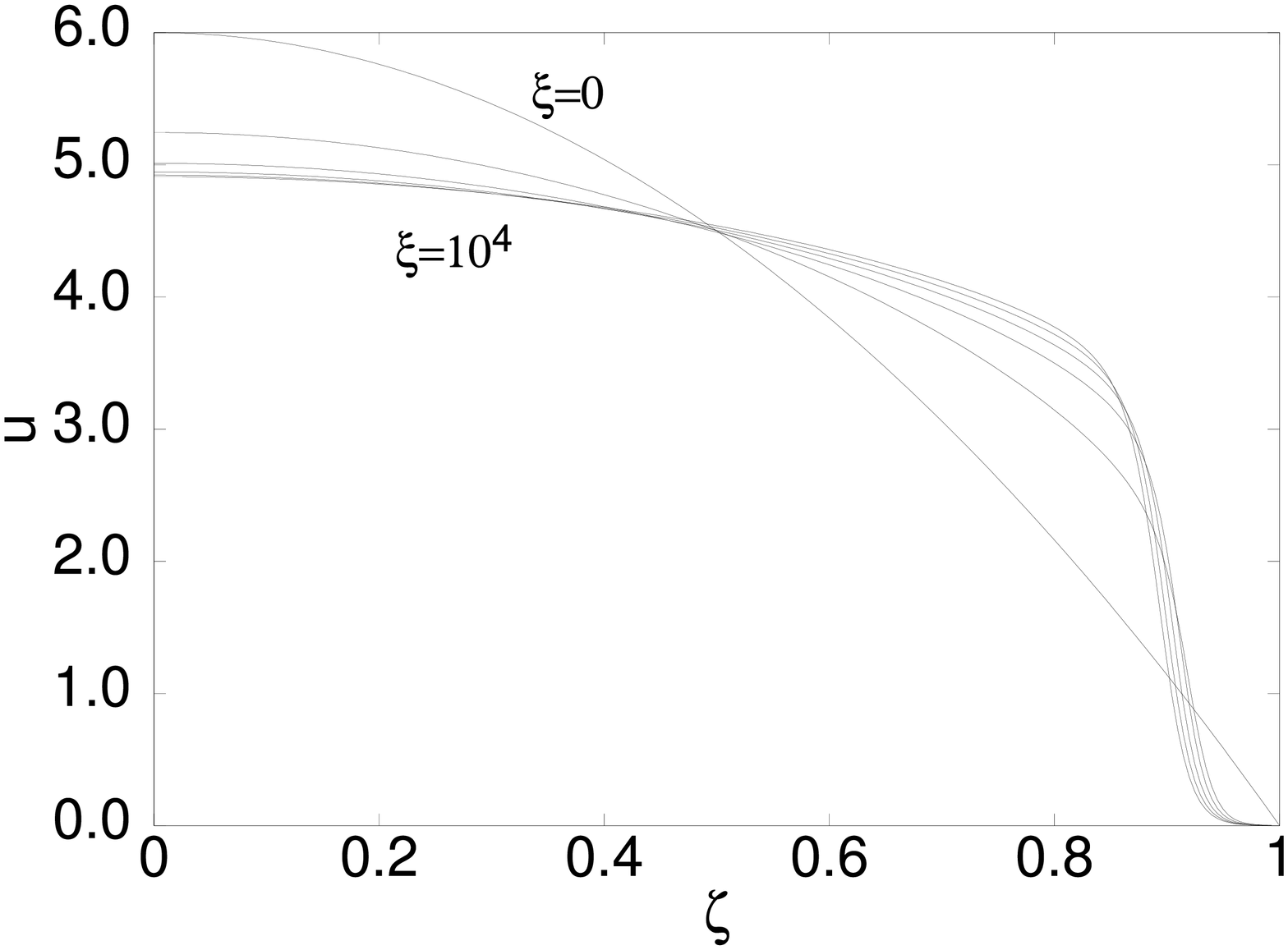}} 
\caption{Longitudinal evolution of the non-dimensional velocity profile
$Pe=10^7$, $Na^*=1000$, $q=4$, ${\mathcal B}=4$ for different non-dimensional
distance from the inlet ($10^{-3}\xi = 0,2,4,6,8,10$).
\label{fig:vel-g1000q4b4}}
\end{figure}

When the viscous heating is important with $\mathcal B >0$,
the typical temperature profile shows low values at the walls with
peaks near the walls, as reported in \figref{fig:temp-g1000q4b4}.
As shown in \figref{fig:vel-g1000q4b4}, starting with a parabolic
velocity profile at the inlet ($\xi=0$), the flow migrates to a
plug-like regime downstream.
By increasing $\mathcal B$, the peak in the temperature
profile moves towards the centre of the channel.
\par
\begin{figure}[htb!]
\figbox*{}{}{%
\includegraphics[angle=-90,width=\figwidth]{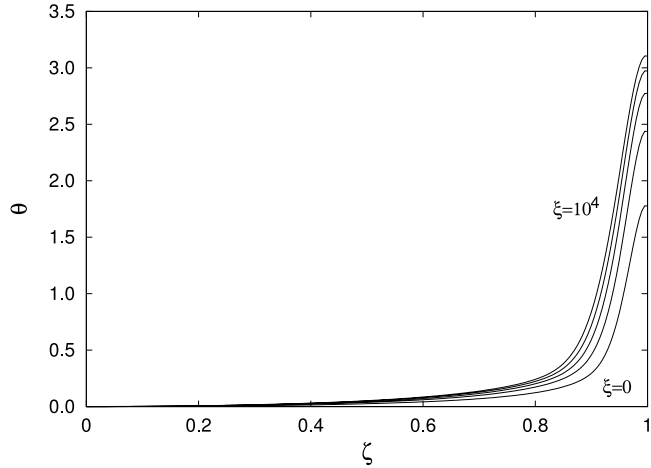}} 
\caption{Longitudinal evolution of the non-dimensional temperature profile
$Pe=10^7$, $Na^*=100$, $q=1$, adiabatic walls, for different non-dimensional
distance from the inlet ($10^{-3}\xi = 0,2,4,6,8,10$).
\label{fig:temp-g100q1ad}}
\end{figure}
\begin{figure}[htb!]
\figbox*{}{}{%
\includegraphics[angle=-90,width=\figwidth]{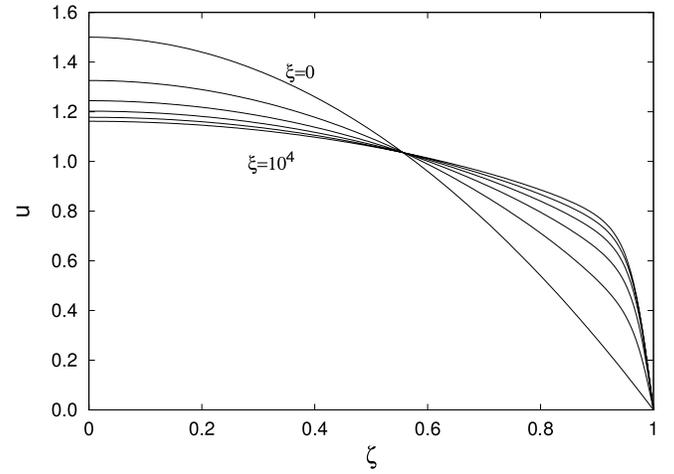}} 
\caption{Longitudinal evolution of the non-dimensional velocity profile
$Pe=10^7$, $Na^*=100$, $q=1$, adiabatic walls, for different non-dimensional
distance from the inlet ($10^{-3}\xi = 0,2,4,6,8,10$).
\label{fig:vel-g100q1ad}}
\end{figure}

In general, viscous heating becomes important when either
the Nahme number ($Na^*$) or the non-dimensional flow rate ($q$)
increase.
For comparison, in \figref{fig:temp-g100q1ad} and
\ref{fig:vel-g100q1ad}, we show the evolution of the temperature and
velocity profiles in a case similar to \figref{fig:temp-g100q1b0} and
\ref{fig:vel-g100q1b0}, but with adiabatic thermal conditions at the
walls. In this case, as expected, the temperature increases more than
the case of isothermal walls, and the the velocity profile
shows a more pronounced plug-like behaviour.
\subsection{Velocity field stability and secondary flows}
The energy equation of fluids with temperature-dependent viscosity in
infinitely long flows is similar to the equation which governs
chemical explosive processes in some materials (described by the
Frank-Kame\-netski equation, [1939]). In this last case,
when the parameter corresponding to ${\mathcal G}$ is greater than its
critical value, ignition occurs. In fluids, however, other
multidimensional degrees of freedom can be activated without the
occurrence of extreme events. For example, in flows between two
parallel planes, cylindrical secondary flows can develop near the
walls, or toroidal secondary flows can occur in circular pipes.
\par
Due to the strong coupling between viscosity and temperature, the
thermal instability generated by viscous heating may trigger an
instability in the velocity field, which cannot be predicted
by a simple isothermal newtonian models.
When the viscous heating produces sharp peak in the temperature profile
near the walls, with consequent strong increase in the 
viscosity gradient, a triggering of instabilities and the transition to
secondary flows (with more efficient thermal and mechanical diffusion)
is possible.
The stability of the plane Couette flow was recently re-examined by 
\citet{yuewen96}, who improve the
results previously found by \citet{sukgol73}.
The flow shows two different instability modes: the
former arises in the non-viscous limit, and the latter is due to the viscosity
stratification \citep{yuewen96}. For this last instability mode, it was
numerically demonstrated that the critical Reynolds number ($Re_c$),
above which the flow becomes turbulent, decreases with the increase of
the Nahme
number ($Na$), that is with the viscous heating. 
This behaviour is confirmed by recent experiments performed by
\citet{whimul2000}. In these
experiments, the authors use a temperature-dependent fluid
(glycerin) and a Taylor-Couette device which allows the tracking of
vortex by a laser particle tracer. Results clearly show that, above a
critical Nahme number, an instability appears at a Reynolds number one
order of magnitude lower than the corresponding Reynolds number
predicted for isothermal flow. 
\par
Moreover, the triggering of the instability in the laminar flow and the
activation of new secondary flows are confirmed by our 2-D direct numerical
simulations of the complete Navier-Stokes equations, which will be the
subject of a further work.
\section{Implications for lava flows}
In this section we apply the previous theoretical results to the study
of lava and magma flows in tubes. We take the typical magma parameters
reported in \tabref{tab:parametri_etna}  as representative, but
the obtained results can be easily generalized.
\begin{table}[htb!]
\caption{Parameters characteristic of lava flows.}
\vspace{0.2cm}
\renewcommand{\tabcolsep}{2pc} % enlarge column spacing
\renewcommand{\arraystretch}{1.2} % enlarge line spacing
\begin{tabular}{ccc}
\hline
$\rho$  & 2500 & kg/m$^3$ \\
$\beta$ & 0.04 & K$^{-1}$ \\
$c$     & 1000 & J\,kg$^{-1}$K$^{-1}$ \\
$k$     & 2.0  & W\,m$^{-1}$K$^{-1}$ \\
$T_s$   & 1173 & K \\
$T_{in}$   & 1373 & K \\
$\mu(T_{in})$ & 500 & Pa\,s\\
$U$     &  1$\div $10  & m/s \\
$2H$     &  0.5$\div$ 5  & m \\
\hline
\end{tabular}
\label{tab:parametri_etna}
\end{table}
\par
Following \citet{bruhup89}, $T_w$ is the temperature at the wall,
which is defined as the temperature at which crystallization of magma
ceases the flow. Here, we estimate $T_w$ as the temperature in the
mid-range between $T_{in}$ and the solidification temperature $T_s$:
therefore, we use $T_0=T_w\approx 1273$K. From this value of $T_w$
and from the values of $\beta$ and $T_{in}$, reported in
\tabref{tab:parametri_etna}, we have ${\mathcal B} =
\beta(T_{in}-T_{w})\approx 4$ and $\mu_0=\mu(T_{in})\cdot e^{\mathcal
B}\approx 30,000$ Pa\,s. Using the other values reported in
\tabref{tab:parametri_etna}, we obtain
the following characteristic non-dimensional numbers:
\be
\begin{array}{ll}
\displaystyle
q=\mu_0Q/\rho g H^3 &\qquad 1\div10\\
Na_0 = \mu_0U^2\beta/k  &\qquad 6\cdot 10^2\div 6\cdot 10^4 \\
\displaystyle
Pe = \rho c U H/k   &\qquad 3\cdot 10^5\div 3\cdot 10^7\\
\displaystyle
Pr_0 = c \mu_0/k      &\qquad 1.5\cdot 10^7  \\
\displaystyle
Na = \mu(T_{in})U^2\beta/k  &\qquad  1\div 10^3 \\
\displaystyle
Re = \rho UH/\mu(T_{in})  &\qquad  1 \div 100 \\
\displaystyle
Pr = c \mu(T_{in})/k      &\qquad 3\cdot 10^5  \\
\end{array}
\label{eq:numeri_adimensionali}
\ee
The last three numbers in \eqref{eq:numeri_adimensionali} are the
Nahme, the Reynolds and the Prandtl numbers based on $\mu(T_{in})$,
respectively, and are reported just for completeness.
\par
The values of the non-dimensional numbers reported in
\paref{eq:numeri_adimensionali} allow for the application of the model 
described above for lava and magma flows. In fact, the high Pecl\'et
number results in a very fast advective transport and validates the
assumption of the lubrication theory.
Concerning the effects of viscous heating on magma flows, we
found that for the lower value of the non-dimensional flow rate
considered in our study ($q = 1$) and $Na^* = Na_0/q^2 \lesssim 1000$, 
the cooling effects prevail on the viscous heating. This becomes
dominant for $Na^* \sim 10^4$ showing velocity and temperature
profiles similar to those reported in \figref{fig:temp-g1000q4b4} and
\ref{fig:vel-g1000q4b4}.
Instead, at high flow rates ($q=10$), the corresponding values of
$Na^*$ are lower: the viscous heating effects are dominant even at
$Na^* \sim 100$ whereas for $Na^* \lesssim 10$ cooling effects prevail.
Since these conditions are usual in magma flows, it should be possible
to observe viscous heating effects in the natural environment.
\par
For magma flows in eruptive conduits, recently,
\citet{polpap2001} emphasized the importance of the viscous heating
in magma flow during volcanic eruptions to explain the
creation and the discharge of two different varieties of pumice. In
their scheme, one type of pumice originates in the region with greater
temperature and higher strain rate near the conduit walls, whereas the other
type is generated in the central part of the conduit with lower strain
rate and temperature.
\par
For lava flows, field evidences corresponding to the starting of
secondary flows, previously described, are possibly represented by
the ``roller vortex'' phenomenon \citep{boosel73} and by the thermal
erosion observed by \citet{grefag98} where the
authors find ``unequivocal evidence for thermal erosion'' in lava
tubes at Cave Basalt, Mt. St.Helens for which the dynamic analysis
indicates laminar flows.
It is known that the turbulent flow is more effective in erosion than the
laminar flow because of the higher heat transfer rate
\citep{hul73,grefag98}; moreover preliminary studies show that purely
thermal erosion by laminar flow is very difficult unless the substrate
is of much lower melting temperature than the eroding
fluid \citep{grefag98}.
Field studies report that during the Kilauea eruption in 1994, lava
eroded 5 meters of the basaltic substrate with an average erosion rate
of 10 cm/day at low Reynolds numbers between 16 and 64 \citep{kaucas98}.
Such high erosion rates highlight the difficulties that arise when
trying to explain thermal erosion in laminar flows. Moreover,
\citet{kaucas98} show that
the Jeffreys equation that relates the velocity and the
flow depth in a laminar flow is less adequate than other relations
valid for turbulent flows such as the equation of Goncharov-Chezy,
even for flows at low Reynolds numbers. Concerning
the temperature, \citet{kaucas98} report the interesting and puzzling
observations of some temperature measurements greater than the magma
temperature at the vent and even greater that the upper limit of the
temperature range of the used radiometer (about 1200~$\deg$C). An
explanation of this phenomenon could be the local increase of
temperature near the walls caused by the viscous heating, as
previously described. 
Other authors also report the evidence of vortices associated with the
beginning of turbulence in active lava flows
\citep{kessel98} and the onset of complex flow patterns
\citep{gre87,lipban87,cribal94,calpin98}.
The inadequacy of pure laminar flows to explain field observation such
as mixing and streamlines breaking has been well known since
the sixties when the not widely accepted term of ``disrupted flow''
was introduced to describe a flow with characteristics between pure
laminar and turbulent \citep[for a discussion see][]{balspu95}.
Moreover basal temperature measurements carried out by \citet{kes95b}
show  that temperature at the base of some flows increases after some
initial  cooling. This behaviour cannot be predicted by a simple
conductive model but can be explained using a correct physical model 
which correctly describes viscous heating effects.  
\par
At Etna, during the eruption of 1971, secondary rotational flows were
observed near the walls in flows confined between levees or in deep
channels \citep{boosel73}. These secondary flows consist of two
elicoidal patterns symmetrical to the centre of the flow. 
\citet{boosel73}, because of the low Reynolds number, exclude any
turbulent flow, and try to explain these rotational flows invoking
other causes and other, sometimes weak, argumentations.
We think that the observed phenomena reported by \citet{boosel73}
could be interpreted as the secondary flows triggered by a pronounced
viscous heating.
The rotational flows, which allow the mixing of 
the fluid along sections orthogonal to the direction of the flow,
invoked in the model of lava flows with two
thermal components by \citet{cribal90}, could have a similar interpretation.
\par
Moreover, exploration and investigations on terrestrial planets and
satellites demonstrated that their lava channels are typically larger
than the corresponding channels on the Earth.
To account for the large sizes of lunar and
martian channels \citet{hul73,hul82} and \citet{car74} proposed that
lavas eroded the ground over which they
flowed (see also \citet{balspu95}). Our model could be used to
investigate whether the action of secondary flows caused by viscous
heating could have eroded the preflow surface more efficiently than
laminar flows.
\par
We wish to remark that the proposed model is applicable to slab flows,
but it easily generalizable to flows in circular pipes.
Moreover, modeling of magma and lava
flows needs also the study of open channels with free surface and
different types of thermal boundary conditions, not considered in the
present study. In particular, a more realistic description should
account for the temperature variations of the ambient medium, that
needs the introduction of additional control parameters such as the
Nusselt number which measures how much the heat transferred through
the boundaries is conducted away.
\par
Finally, an accurate study of the flow instabilities and
the triggering of secondary flows needs the solution of the complete
transport equations in 2-D or 3-D, which is the subject of a further work in
preparation.
\section{Conclusions}
The effects of viscous heat generation in fluids with strongly
temperature-dependent viscosity such as silicate melts are
investigated. These effects can play an important role in the
dynamics of magma flow in conduits and lava flows in tubes.
In fact, viscous friction generates a local increase in temperature
near the tube walls, with consequent increase of the fluid velocity
because of the viscosity decrease.
In the case of infinitely long tubes, one-dimensional models predict
that viscous heating can lead to a positive feed-back known as
``thermal runaway''. Actually, as described by 2-D and 3-D models,
this feed-back causes the activation of other degrees of
freedom, with the production of local instabilities and the triggering
of secondary flows.
However, the typical high value of the Pecl\'et number in magma flows
does not allow the assumption of an infinitely long tube (from the
thermal point of view), and a model for magma flow in tubes of finite
length is needed, as described in the paper. 
In general, this process is controlled by the value of the Nahme
number, the flow rate, the Pecl\'et number, and the thermal conditions
at the inlet. 
By adopting such a model, and typical values of the parameters for
lava flows in tubes, we find that starting from a constant temperature
and parabolic velocity profile at the tube inlet, viscous heating
causes the increase of the temperature near the walls, with a
consequent local viscosity decrease. 
This can lead to the formation of a plug-like velocity profile,
commonly observed in lava flows, and explained only by assuming a
Binghamian rheology. Moreover, the presence of an inflex in the
velocity profile near the walls can lead to the formation of
local flow instabilities, even at low Reynolds numbers.
The assumption of isothermal boundary conditions at the walls,
although simplifying, is probably too restrictive; however, by
assuming adiabatic conditions, the qualitative character of the flow
does not seem to change dramatically in tubes of finite length.
We hope that this work will provide a motivation for further
theoretical and field investigations of viscous heating effects in
magma flows. In particular, the formation of vortices in active lava
flows at low Reynolds numbers, thermal erosion, temperature and
velocity profile. Moreover, the effects viscous dissipation in
volcanic conduits could have an important role on the dynamics of the
both effusive and explosive eruptions.
\begin{acknowledgements}
This work was supported by the European Commission (Contract
ENV4-CT98-0713), and by the Gruppo Nazionale per la Vulcanologia
(INGV). We are grateful to all the researchers who contributed to
improve the manuscript with their helpful suggestions, and in
particular to A.~Neri. Finally we wish to thank S.~Baloga and the 
anonymous referees of this paper.
\end{acknowledgements}
%
% *************** References ***************
%
\bibliography{references}
\end{document}